\documentclass[pra,twocolumn,tightenlines,showpacs,nofootinbib]{revtex4}
\usepackage{bm,dcolumn,amsmath,graphicx}

\newcommand{\eref}[1]{Eq.~(\ref{#1})}
\newcommand{\tref}[1]{Table~\ref{#1}}

\begin{document}

\title{Effect of atomic electrons on 7.6 eV nuclear transition
 in $^{229}$Th$^{3+}$}

\author{S. G. Porsev$^{1,2}$}
\author{V. V. Flambaum$^1$}
\affiliation{$^1$ School of Physics, University of New South Wales,
Sydney, NSW 2052, Australia}
\affiliation{$^2$ Petersburg Nuclear Physics Institute, Gatchina,
Leningrad district, 188300, Russia}

\date{ \today }

\begin{abstract}
We have considered an effect of atomic electrons due to
the electronic bridge process on the nuclear
$^{229m}$Th -- $^{229g}$Th transition in $^{229}$Th$^{3+}$. Based on
a recent experimental result we assumed the energy difference
between the isomeric and the ground nuclear states to be equal to 7.6 eV.
We have calculated the ratios of the electronic bridge process probability
($\Gamma_{\rm EB}$) to the probability of the nuclear radiative transition
($\Gamma_N$) for the electronic $5f_{5/2} \rightarrow 6d_{3/2},6d_{5/2},7s$ and the
$7s \rightarrow 7p_{1/2},7p_{3/2}$ transitions and found
$\Gamma_{\rm EB}/\Gamma_N \sim 0.01\div0.1$ for the former and
$\Gamma_{\rm EB}/\Gamma_N \sim 20$ for the latter.
\end{abstract}

\maketitle

\section{Introduction}
\label{sec_I}
The $^{229}$Th nucleus is unique in a sense that the energy
splitting of the ground state doublet is only several eV. Though
a prediction of existence of so low-lying level has
been made more than thirty years ago~\cite{KroRei76},
the definite value of energy of the isomeric state $^{229m}$Th
is not known so far. In 1990 Reich and Helmer~\cite{ReiHel90}
measured this excitation energy ($\omega_N$) to be $3.5 \pm 1.0$ eV.
In Ref.~\cite{GuiHel05} it was obtained $5.5 \pm 1.0$ eV.
Finally, a recent experiment of Beck {\it et al.}~\cite{BecBecBei07}
has given even larger value (with least error)
$\omega_N = 7.6 \pm 0.5$ eV.

As to the lifetime of the $^{229m}$Th, measurements performed by different experimental groups
led to different values. The results differ from each other by several orders of magnitude,
changing from a few minutes~\cite{InaHab09}
to many hours~\cite{MitHarOht03}. Hence, new experimental and theoretical
investigations are required.

A special interest to the nuclear transition from the isomeric state to the
ground state is motivated by a possibility to build a
superprecise nuclear clock~\cite{PeiTam03} and very high sensitivity to the effects of
possible temporal variation of the fundamental constants including the fine structure
constant $\alpha$, strong interaction and quark mass~\cite{Fla06}.

Laser cooling of the $^{232}$Th$^{3+}$ ion
was recently reported by Campbell {\it et al.} in their paper~\cite{CamSteChu09}.
This was the first time when a multiply charged ion has
been laser cooled. As a next step this experimental group plans
to investigate the nuclear transition between the isomeric and the ground state
in a trapped, cold $^{229}$Th$^{3+}$ ion.
Motivated by this experimental progress we have
considered $^{229}$Th$^{3+}$ ion and
calculated the transition probability of the $^{229}$Th nucleus
from its lowest-energy isomeric state $^{229m}$Th to the ground state  $^{229g}$Th due to
the electronic bridge (EB) process.

Our calculations, based on the value of $\omega_N$ = 7.6 eV,
showed that if the electrons are in their ground state
the ratio of the probability of the EB process, $\Gamma_{\rm EB}$,
to the probability of the nuclear radiative $M1$ transition, $\Gamma_N(M1)$,
is of the order of (a few)$\times 10^{-2}$. If the valence electron is in the metastable $7s$ state
then $\Gamma_{\rm EB}/\Gamma_N(M1) \sim 20$.

The paper is organized as follows.
In Section~\ref{sec_GF} we present the general formalism describing
the EB process. Section~\ref{sec_MC} is devoted to the method of calculation
of the properties of Th$^{3+}$. In Section~\ref{sec_RD} we discuss the results
of calculations and Section~\ref{sec_C} contains concluding remarks. If not stated
otherwise the atomic units ($\hbar = |e| = m_e = 1$ and the speed of light $c = 137$)
are used.
\section{General formalism}
\label{sec_GF}
\subsection{Configurations mixing between combined electron-nucleus states}
The 7.6 eV transition in $^{229}$Th is the M1 transition with the amplitude
of a fraction of the nuclear magneton $\mu_N$. An amplitude of an allowed
electric dipole transition of the valence electron, $\sim$ 1 au,
is $10^6$ times larger. If there is an electron excited state close to the
energy of the nuclear excitation, an energy transfer from the nuclear excited state to
the electron excited state accompanied by the electron electric dipole transition to a lower
state, may radically decrease the lifetime of the nuclear isomeric state.
Even if there is no an electron state very close to the nuclear excited state,
the electron bridge process produces significant effect.

\begin{figure}
\includegraphics[scale=0.5]{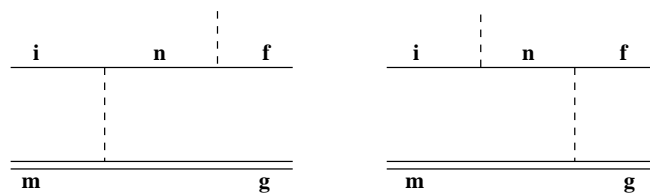}
\caption{The single and double solid lines relate to the electronic and the nuclear
transitions, correspondingly. The dashed line is the photon line.}
\label{Fig:EB}
\end{figure}

The EB process can be represented by two Feynman diagrams in Fig.~\ref{Fig:EB}.
In the following we assume that the initial $i$ and the final $f$ electronic states are of
opposite parity and a real photon which is emitted or absorbed is the electric dipole photon.
The probability of the EB process in this case is much larger than
in the case when the $i$ and the $f$ states are of the same parity.

Therefore, the EB process can be effectively treated as the electric
dipole $i \rightarrow f$ transition of the electron accompanied by the nuclear
transition from its isomeric
state to the ground state. Denoting by ${\bf D}_{\rm EB}$ the amplitude of this
``generalized'' electric dipole transition and assuming that the initial and
the final states are fixed, we obtain
\begin{eqnarray}
 {\bf D}_{\rm EB} &=& \sum_n
\frac{\langle f|{\bf D}|n \rangle \langle g,n|H_{\rm int}|m,i \rangle}
{\varepsilon_i + E_m - \varepsilon_n - E_g + i\Gamma_n/2} + \nonumber \\
&&\sum_{k}
\frac{\langle g,f |H_{\rm int}| m,k \rangle \langle k|{\bf D}|i \rangle}
{\varepsilon_f + E_g - \varepsilon_k - E_m + i\Gamma_k/2} ,
\label{D_EB}
\end{eqnarray}
where the indices $i$, $(n,k)$, and $f$ denote initial, intermediate, and final
electronic states, correspondingly; and the indices
$g$ and $m$ denote the ground state and the isomeric state of the nucleus.
$\varepsilon_l$ are the atomic energies, $E_{m,(g)}$ are the nuclear energies
of the isomeric (ground) states, and $\Gamma_l$ are the widths of the intermediate
states which may be neglected in the case of $^{229}$Th.
The operator ${\bf D}=-{\bf r}$ is the electron electric dipole moment operator and
$H_{\rm int}$ is the hyperfine coupling Hamiltonian, which may be represented as a
sum over multipole nuclear moments $\mathcal{M}_{K}^\lambda$ of rank $K$ combined
with the even-parity electronic coupling operators $\mathcal{T}_{K \lambda}$
of the same rank as
\begin{equation}
H_{\rm int}= \sum_{K \lambda} \mathcal{M}_K^\lambda \mathcal{T}_{K \lambda} .
\label{H_int}
\end{equation}

Neglecting the hyperfine splitting of levels, we can represent the total
wave function as a product of the nuclear wave function and the electronic
wave function. For instance,
$|g,n\rangle = |g\rangle |n\rangle \equiv |I_g M_g\rangle |\gamma_{n}J_{n}m_{n}\rangle$,
where $I_g$ is the nuclear spin, $M_g$ is the projection of the nuclear spin;
$J_n$ is the electron total angular momentum, $m_n$ is its projection, and
$\gamma_n$ encapsulates all other electronic quantum numbers. Taking into account
\eref{H_int} we can rewrite~\eref{D_EB} as
\begin{eqnarray}
{\bf D}_{\rm EB} &=& \sum_{K \lambda} \left[  \sum_{n}%
\frac{\langle f|{\bf D}|n\rangle\langle n|\mathcal{T}_{K \lambda}|i\rangle}
{\omega_{in}+\omega_{N}} \right. \nonumber \\
&+& \left. \sum_{k}\frac{\langle f|\mathcal{T}_{K \lambda}|k\rangle
\langle k|{\bf D}|i\rangle} {\omega_{fk}-\omega_{N}} \right]
\langle g|\mathcal{M}_K^\lambda|m \rangle ,
\label{D}%
\end{eqnarray}
where $\omega_{ab} \equiv \varepsilon_{a}-\varepsilon_{b}$ and
$\omega_N = E_m - E_g$.

Thus, we need to carry out the atomic calculation which is
similar to that for a forbidden $E1$ transition opened by the
hyperfine interaction (see, e.g.,~\cite{PorDer04}). The only difference is that the matrix element (ME)
of the nuclear moment $\langle g|\mathcal{M}_K^\lambda|m \rangle$ here is non-diagonal
(there is also a few per cent correction due to variation of the electron
magnetic field inside the nucleus). Note that the conventionally defined nuclear
moments are related to the tensors $\mathcal{M}_{K}^\lambda$ as
$\mu \equiv \langle I M_I=I| \mathcal{M}_1^0 |I M_I=I \rangle$ and
$Q \equiv 2 \langle I M_I=I| \mathcal{M}_2^0 |I M_I=I \rangle$.

The probability $\Gamma_{\rm EB}$ of the electric dipole transition
determined by its amplitude ${\bf D}_{\rm EB}$ is given by a simple
formula (see, e.g., \cite{BerLifPit82})
\begin{equation}
\Gamma_{\rm EB} =\frac{4}{3}\left(  \frac{\omega}{c}\right)^3
|\mathbf{D}_{\mathrm{EB}}|^2 ,
\label{GamEB_1}
\end{equation}
where $\omega$ is the real photon frequency determined from the low of conservation of energy
as $\omega = \varepsilon_i - \varepsilon_f + \omega_N$.

If we average over the initial projections of the electronic and the nuclear total angular momenta
$m_{i}$ and $M_{m}$ and summing over the final projections $m_{f}$ and $M_{g}$,
\eref{GamEB_1} is transformed to
\begin{eqnarray}
\Gamma_{\rm EB} &=&\frac{4}{3}\left(  \frac{\omega}{c}\right)^{3}
\frac{1}{\left(2I_{m}+1\right)  \left(  2J_{i}+1\right)  } \times \nonumber \\
&&\underset{M_{m}M_{g},m_{i}m_{f}}{%
{\displaystyle\sum}
}\left\vert \mathbf{D}_{\mathrm{EB}}\right\vert ^{2} .
\label{GamEB_2}
\end{eqnarray}
Substituting \eref{D} to \eref{GamEB_2}, applying the Wigner-Eckart theorem
and performing the summation over all magnetic quantum numbers of the initial,
intermediate and final states, we can reduce \eref{GamEB_2} to the
form $\Gamma_{\rm EB} = \sum_K \Gamma^{(K)}_{\rm EB}$,
where $\Gamma^{(K)}_{\rm EB}$ can be represented by
\begin{eqnarray}
\Gamma^{(K)}_{\rm EB} &=& \frac{4}{3}\left( \frac{\omega}{c}\right)^3
\frac{|\langle I_g||\mathcal{M}_K||I_m \rangle|^2}{(2K+1)(2I_m+1)(2J_i+1)} \nonumber \\
&\times& \left( G_1^{(K)} + G_{12}^{(K)} + G_2^{(K)} \right) ,
\label{GamK}
\end{eqnarray}
where
\begin{eqnarray}
 G_1^{(K)} &\equiv& \sum_{J_n} \frac{1}{2J_n+1}  \nonumber \\
&\times& \left\vert \sum_{\gamma_k} \frac
{\langle \gamma_f J_f||D||\gamma_k J_n \rangle
 \langle \gamma_k J_n||\mathcal{T}_K|| \gamma_i J_i \rangle}
{\omega_{ik}+\omega_N} \right\vert ^2,
\label{G1}
\end{eqnarray}
\begin{eqnarray}
 G_{12}^{(K)} &\equiv& 2 \sum_{J_t J_n} (-1)^{J_t+J_n}
\left\{
\begin{array}
[c]{ccc}%
J_i & J_t & 1\\
J_f & J_n & K
\end{array}
\right\}  \nonumber \\
&\times& \sum_{\gamma_k}
\frac{\langle \gamma_f J_f ||D|| \gamma_k J_n \rangle
      \langle \gamma_k J_n ||\mathcal{T}_K|| \gamma_i J_i \rangle}
     {\omega_{ik}+\omega_N}   \nonumber \\
&\times& \sum_{\gamma_s}
\frac{\langle \gamma_f J_f ||\mathcal{T}_K|| \gamma_s J_t \rangle
      \langle \gamma_s J_t ||D|| \gamma_i J_i \rangle}
     {\omega_{fs}-\omega_N} ,
\label{G12}
\end{eqnarray}
and
\begin{eqnarray}
G_2^{(K)} &\equiv& \sum_{J_n} \frac{1}{2J_n+1}  \nonumber \\
&\times& \left\vert \sum_{\gamma_k}
\frac{\langle \gamma_f J_f ||\mathcal{T}_K|| \gamma_k J_n \rangle
      \langle \gamma_k J_n ||D|| \gamma_i J_i \rangle}
     {\omega_{fk}-\omega_N}\right\vert^2\!.
\label{G2}
\end{eqnarray}
The terms $G_1^{(K)}$ and $G_2^{(K)}$ characterize the contributions
of the first and second diagrams in Fig.~\ref{Fig:EB} while the ``interference''
of two these diagrams is given by $G_{12}^{(K)}$.

It is worth noting that \eref{GamK} is valid in a general case
because deriving it we did not make any approximations.
In particular, we did not suppose
that there is an electronic transition whose frequency is close to the
the nuclear transition frequency $\omega_N$.
In systems where such a ``resonance'' transition exists,
the expression for $\Gamma_{\rm EB}$ can be significantly simplified.
\subsection{Derivation of the coefficients $\beta_{M1}$ and $\beta_{E2}$}
Since the frequency of the nuclear transition
from the isomeric state to the ground state of $^{229}$Th is very small,
in the following we will take into consideration only first two terms in~\eref{GamK},
involving the nuclear magnetic-dipole ($K=1$) and electric-quadrupole
($K=2$) moments.
Another consequence of the smallness of the nuclear transition frequency is that
the probability of the $m \stackrel{E2}{\longrightarrow} g$ transition $\Gamma_N(E2)$ is
strongly suppressed in comparison to the probability of
$m \stackrel{M1}{\longrightarrow} g$ transition $\Gamma_N(M1)$.

The probability $\Gamma_N(\tau K, m \rightarrow g)$ of the $\tau K$ transition
(where $\tau$ denotes $M$ or $E$) in the $^{229}$Th nucleus can be written
in a form used in the nuclear physics as (see, e.g.,~\cite{Seg77})
\begin{eqnarray}
&&\Gamma_N (\tau K, m \rightarrow g) = \nonumber \\
&& 8 \pi \frac{k_N^{2K+1}}{[(2K+1)!!]^2} \, \frac{K+1}{K}\,\, B(\tau K, m \rightarrow g) .
\label{Gam_N}
\end{eqnarray}
Here $k_N \equiv \omega_N/c$ and the reduced probability of the nuclear $m \rightarrow g$ transition
$B(\tau K, m \rightarrow g)$, expressed in terms of the operator $\mathcal{M}_K$, reads as
\begin{eqnarray}
B(\tau K, m \rightarrow g) = 
\frac{1}{2I_m+1} \frac{2K+1}{4 \pi}
|\langle I_g || \mathcal{M}_K || I_m \rangle|^2 .
\label{B}
\end{eqnarray}
Using \eref{Gam_N} we find for this transition
\begin{eqnarray}
\frac{\Gamma_N(M1)}{\Gamma_N(E2)} &=& \frac{100}{3} \frac{1}{k_N^2} \,
\frac{B(M1, m \rightarrow g)}{B(E2, m \rightarrow g)} .
\label{GM1_GE2}
\end{eqnarray}
The theoretical value of $B(M1, m \rightarrow g)$ was obtained in~\cite{DykTka98}
\begin{equation}
B(M1, m \rightarrow g) \approx 0.086\, \mu_N^2 .
\label{B_M1}
\end{equation}
To the best of our knowledge the accurate value of $B(E2, m \rightarrow g)$ is unknown.
An estimate of this quantity is found in Ref.~\cite{StrTka91}, where
Strizhov and Tkalya, referring to the paper~\cite{BemMcGPor88}, cite the value of
several Weisskopf units (W.u.) for $B(E2, m \rightarrow g)$.

The definition of 1 W.u. for the $E2$ transition from a nuclear excited state to the
ground state (in usual units) is
\begin{equation}
E2\!: \,\, 1\, {\rm W.u.} = 5.940 \times 10^{-6} A^{4/3}\, (e \cdot {\rm barn})^2 ,
\label{Wu}
\end{equation}
where $e$ is the electron charge and $A$ is the number of nucleons in the nucleus.

Using Eqs.~(\ref{GM1_GE2}), (\ref{B_M1}), and~(\ref{Wu}) we arrive at the estimate
\begin{equation}
\frac{\Gamma_N(M1)}{\Gamma_N(E2)} \sim 10^{11} .
\label{M1E2}
\end{equation}
An accurate calculation of the probability of the nuclear
$E2$ transition is beyond the topic of this work. In the following
we rely on the estimate given by~\eref{M1E2} and
concentrate our efforts on the computation of the ratios
$$\beta_{M1} = \Gamma_{\rm EB}^{(1)}/\Gamma_N(M1) \quad {\rm and} \quad
 \beta_{E2} = \Gamma_{\rm EB}^{(2)}/\Gamma_N(E2),$$
where $\Gamma_{\rm EB}^{(1,2)}$ are given by~\eref{GamK} and $\Gamma_N(M1)$
and $\Gamma_N(E2)$ can be found from Eq.~(\ref{Gam_N}).

Using these equations we obtain
\begin{eqnarray}
\beta_{M1} &=& \left( \frac{\omega}{\omega_N}\right)^3
\frac{1}{3 (2J_i+1)} \nonumber \\
&\times& \left( G_1^{(1)} + G_{12}^{(1)} + G_2^{(1)} \right)
\label{betaM1}
\end{eqnarray}
and
\begin{eqnarray}
\beta_{E2} &=& \left( \frac{\omega}{\omega_N}\right)^3
\frac{1}{k_N^2} \, \frac{4}{(2J_i+1)} \nonumber \\
&\times& \left( G_1^{(2)} + G_{12}^{(2)} + G_2^{(2)} \right) .
\label{betaE2}
\end{eqnarray}

As follows from the estimate \eref{M1E2}, the probability of the nuclear radiative $E2$
transition from the isomeric state to the ground state in $^{229}$Th is completely
negligible in comparison with the probability of the $M1$ transition.
Based on this estimate one can expect that the electronic part of the EB
process mainly contributing to $\Gamma_{\rm EB}$ can be represented as
$i \stackrel{\mathcal{T}_1}{\longrightarrow} n \stackrel{E1}{\longrightarrow} f$,
while the channel
$i \stackrel{\mathcal{T}_2}{\longrightarrow} n \stackrel{E1}{\longrightarrow} f$
can be neglected.

As we will demonstrate below this assumption is valid for
$^{229}$Th$^{3+}$ in spite of that $\beta_{E2}$ is
many orders of magnitude larger than $\beta_{M1}$.
The physical meaning of this is as follows. It is known that a neutral atom is not affected by an
external electric field. It means that an effective electric field acting
on the nucleus is equal to zero because the electrons completely
screen the external electric field. Respectively, gradient of electrostatic
potential created by the electrons at the nucleus is very large. For a
static case (in our consideration it corresponds to $\omega_N =0$) a similar
phenomenon was investigated in~\cite{FeiJoh69,KolJohSho82} where
magnetic-dipole shielding factors and electric-quadrupole antishielding factors
were calculated for a number of atoms and ions. For instance, for such a
heavy atom as Hg, the latter was shown to be four orders of magnitude
larger than the former.

Note also that the probability of, so called, ``elastic'' process (when the final
state is the same as the initial state) is much smaller, since
instead of $E1$ transitions we have to consider $M1$ (or $E2$) transitions.
But the probability of an allowed $M1$ transition is four orders of magnitude smaller than
the probability of an allowed $E1$ transition.

The triply ionized thorium $^{229}$Th$^{3+}$ is an univalent ion. Respectively,
the total electronic angular momentum as
well as other quantum numbers coincide with the quantum numbers of the
valence electron. The expressions for the single-electron operators $T_1$ and $T_2$
and for the MEs of the operators $D$, $T_1$ and $T_2$ are presented
in the Appendix~\ref{Ap}.

\section{Method of calculation}
\label{sec_MC}
At the first stage we have solved Dirac-Hartree-Fock (DHF)
equations~\cite{BraDeyTup77} in $V^{N-1}$ approximation. It means
that the DHF equations were solved self-consistently for the core
electrons. After that we determined valence orbitals for several
low-lying states from the frozen-core DHF equations. The virtual
orbitals were determined with the help of a recurrent
procedure~\cite{KozPorFla96}. One-electron basis set of the
following size was constructed:
$ 1-20s, \, 2-20p, \, 3-20d, \, 4-25f, \, 5-18g.$

To find wave functions needed for calculation of $\beta_{M1}$
and $\beta_{E2}$ we applied a relativistic many-body method
initially suggested in Refs.~\cite{DzuFlaSil87,DzuFlaKoz96b} and
subsequently developed in~\cite{DzuKozPor98,PorRakKoz99P}. In this
method one determines wave functions from solution of the
effective many-body Schr\"{o}dinger equation
\begin{equation}
H_{\rm eff}(E_n) \, | \Psi_n \rangle = E_n \, |\Psi_n \rangle \, ,
\label{H}
\end{equation}
with the effective Hamiltonian defined as
\begin{equation}
H_{\rm eff}(E) = H_\mathrm{FC} + \Sigma(E) \, .
\label{Heff}
\end{equation}
Here $H_\mathrm{FC}$ is the frozen-core DHF
Hamiltonian and self-energy operator $\Sigma$ is the
energy-dependent correction, involving core excitations,
which recovers second order of perturbation theory
in residual Coulomb interaction and additionally accounts for
certain classes of many-body diagrams in all orders of
perturbation theory. We will refer to this approach as the DHF+$\Sigma$
formalism.

Together with the effective Hamiltonian $H_{\rm eff}$ we
introduce effective (``dressed'') electric-dipole operator $D_{\rm eff}$ and
operators $(\mathcal{T}_K)_{\rm eff}$ acting in the model
space of valence electrons. These operators were obtained within
the  relativistic random-phase approximation
(RPA)~\cite{DzuKozPor98,KolJohSho82} which
describes a shielding of the externally applied
electric field by the core electrons. The RPA sequence
of diagrams was summed to all orders of the perturbation theory.

\begin{figure}
\includegraphics[scale=0.5]{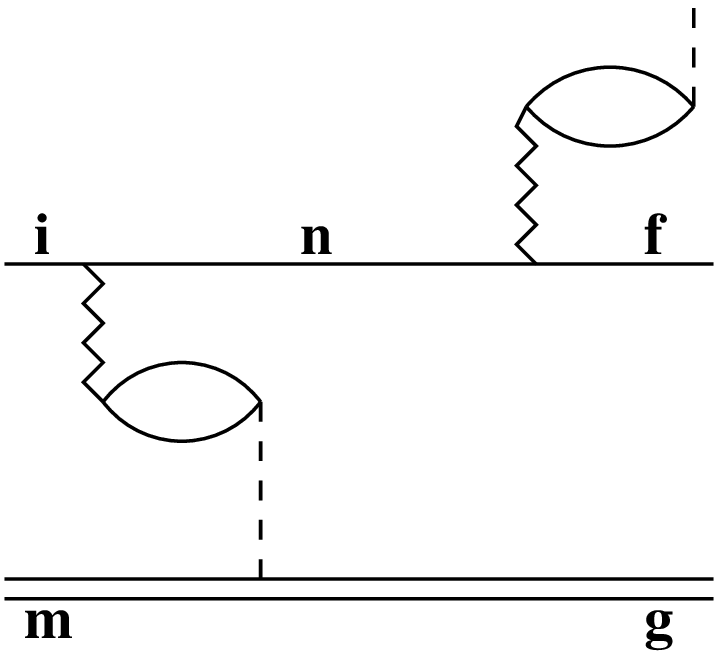}
\caption{The single and double solid lines relate to the electronic and the nuclear
transitions, correspondingly. The dashed line is the photon line. The wavy
line is the Coulomb interaction between electrons.}
\label{Fig:RPA}
\end{figure}

A representative diagram illustrating a contribution of the
RPA corrections in the first order is shown in Fig.~\ref{Fig:RPA}.
As we will show below in certain cases including the RPA corrections
is very important because it changes $\Gamma_{\rm EB}$ by orders
of magnitude.

With the wave functions obtained from Eq.~(\ref{H}), the
quantities $G_1^{(K)}$, $G_{12}^{(K)}$, and $G_2^{(K)}$ can be computed with the
Sternheimer~\cite{Ste50} or Dalgarno-Lewis \cite{DalLew55} method
implemented in the DHF+RPA+$\Sigma$ framework.

For instance, the expression for $G_2^{(K)}$, given by \eref{G2}, can be
rewritten as
\begin{eqnarray}
G_2^{(K)} = \sum_{J_n} \frac{1}{2J_n+1}
\left\vert
\langle \gamma_f J_f ||\mathcal{T}_K|| \delta \psi, J_n \rangle \right\vert^2 ,
\label{G2new}
\end{eqnarray}
where an intermediate-state wave function $|\delta \psi \rangle$
can be found from the inhomogeneous equation
\begin{eqnarray}
|\delta \psi \rangle =
 \frac{1}
{\varepsilon_f - \omega_N - H_{\rm eff}} \, D_z | i \rangle
\label{psif}
\end{eqnarray}
and then $|\delta \psi, J_n \rangle$ is obtained by projecting the
wave function $|\delta \psi \rangle$ to the state with the definite
value of $J_n$.
Similarly we can derive the expressions for $G_1^{(K)}$ and $G_{12}^{(K)}$.

Only excitations of the valence electron to
higher virtual orbitals are included in the intermediate-state wave function
$|\delta \psi \rangle$ due to the presence of $H_{\mathrm{eff}}$ in Eq.~(\ref{psif}).
Additional contributions to $G_1^{(K)}$, $G_{12}^{(K)}$, and $G_2^{(K)}$
come from particle-hole excitations of the core. The role of these
contributions will be discussed more detailed in the next section.

Since Th$^{3+}$ is an univalent element, the quantities $G_1^{(K)}$, $G_{12}^{(K)}$,
and $G_2^{(K)}$ can be obtained by another method. We can directly
sum over {\it all} intermediate states using the single-electron wave functions
found at the stage of constructing the basis set. An accuracy of this approach is
comparable to the accuracy of the more refined method of solving the inhomogeneous
equation. The reason is that, despite a non-resonant character of the EB process
in $^{229}$Th$^{3+}$ for $\omega_N = 7.6$ eV there are only a few intermediate states in
Eqs.~(\ref{G1}), (\ref{G12}), and (\ref{G2}) (whose denominators are small)
that give a dominant contribution to $\Gamma_{\rm EB}$.

We would like to stress that in the sums over the intermediate states in
Eqs.~(\ref{G1}), (\ref{G12}), and (\ref{G2}) the states
$(\gamma_i J_i) = (\gamma_k J_n)$ or
$(\gamma_f J_f) = (\gamma_k J_n)$ are allowed. This is due to that the ME of the nuclear
moment $\langle g|\mathcal{M}_K^\lambda|m \rangle$ is non-diagonal and,
correspondingly, the diagonal MEs of the electronic operator $T_K$
should be included into consideration. Note that the diagonal MEs of the operators
$T_K$ are large and the inclusion of these contributions to $\Gamma_{\rm EB}$
significantly affects the final value of the latter.

\section{Results and discussion}
\label{sec_RD}
To check the quality of the constructed wave functions we have calculated
the energy levels for a number of low-lying states and compared them with
the experimental data. Some details regarding the energy levels computation
can be found in our recent paper~\cite{FlaPor09}. We present in~\tref{Tab:E}
the results obtained on the stage of pure DHF approximation and in the frame
of DHF+$\Sigma$ formalism.

As seen from~\tref{Tab:E} on the  stage of the DHF approximation the order of the low-lying
levels is incorrect. For instance, the $6d_{3/2}$ state lays deeper than
the $5f_{5/2}$ state. An agreement between theoretical and experimental
energy levels is rather poor. The inclusion of the core-valence correlations
restores the correct order of the states and significantly improves the agreement
with the experimental energy levels. Nevertheless in certain cases (e.g., for the $7s$
state) the energy levels were reproduced not very accurately. For this reason in the following calculation
of $\beta_{M1}$ and $\beta_{E2}$
we used the experimental energies for the low-lying states.

\begin{table}
\caption{The low-lying energy levels (in cm$^{-1}$) in the DHF and
the DHF+$\Sigma$ approximations are presented.
The theoretical values are compared with the experimental data.}

\label{Tab:E}

\begin{ruledtabular}
\begin{tabular}{lclcc}
\multicolumn{2}{c}{DHF} &\multicolumn{2}{c}{DHF+$\Sigma$}
                        &\multicolumn{1}{c}{Experiment\footnotemark[1]} \\
\hline
$6d_{3/2}$ &     --- &   $5f_{5/2}$\footnotemark[2]
                                    &    ---    &     ---   \\
$6d_{5/2}$ &    4225 &   $5f_{7/2}$ &    4798   &     4325  \\
$5f_{5/2}$ &    5190 &   $6d_{3/2}$ &    9091   &     9193  \\
$5f_{7/2}$ &    8617 &   $6d_{5/2}$ &   14835   &    14486  \\
$7s_{1/2}$ &   11519 &   $7s_{1/2}$ &   21321   &    23131  \\
$7p_{1/2}$ &   46702 &   $7p_{1/2}$ &   59436   &    60239  \\
$7p_{3/2}$ &   58225 &   $7p_{3/2}$ &   72677   &    73056  \\
$8s_{1/2}$ &  102595 &   $8s_{1/2}$ &  120085   &   119622  \\
$7d_{3/2}$ &  103148 &   $7d_{3/2}$ &  120898   &   119685  \\
$7d_{5/2}$ &  104763 &   $7d_{5/2}$ &  122657   &   121427  \\
$6f_{5/2}$ &  111874 &   $6f_{5/2}$ &  128734   &   127262  \\
$6f_{7/2}$ &  112316 &   $6f_{7/2}$ &  129227   &   127815  \\
$8p_{1/2}$ &  117185 &   $8p_{1/2}$ &  135144   &   134517  \\
$8p_{3/2}$ &  122194 &   $8p_{3/2}$ &  140558   &   139871  \\
$9s_{1/2}$ &  142328 &   $9s_{1/2}$ &  161481   &   160728  \\
\end{tabular}
\end{ruledtabular}
\footnotemark[1]{Reference~\cite{NIST}}; \\
\footnotemark[2]{The removal energy of the $5f_{5/2}$ state was found to
be equal to 0.9414 au on the DHF stage and 1.0584 au on the (DHF+$\Sigma$) stage.
The experimental value is 1.0588 au}.
\end{table}

We have carried out calculations of the coefficients $\beta_{M1}$ and $\beta_{E2}$
for $\omega_N = 7.6$ eV considering the ground state $5f_{5/2}$ and the metastable state $7s$
as the initial state $i$. As follows from the discussion above
the final states should be of opposite parity
in comparison to the initial states. Respectively, we considered
$6d_{3/2}$, $6d_{5/2}$ and $7s$ states to be the final states when the initial
state was $5f_{5/2}$. The $7p_{1/2}$ and the $7p_{3/2}$ states were the final states when
the initial state was $7s$.

In \tref{betaEB} we present the results obtained
1) on the stage of pure DHF approximation, 2) in the DHF + RPA approximation,
and 3) in the frame of DHF + RPA + $\Sigma$ formalism.
\begin{table}
\caption{The coefficients $\beta_{M1}$
obtained for certain $i \rightarrow f$ transitions for $\omega_N = 7.6$ eV
in the DHF, the DHF+RPA, and the DHF+RPA+$\Sigma$ (denoted as $+\Sigma$) approximations
are presented. The coefficients $\beta_{E2}$ are given in the DHF+RPA approximation.}

\label{betaEB}

\begin{ruledtabular}
\begin{tabular}{lccccc}
           &            &\multicolumn{3}{c}{$\beta_{M1}$}
                                                                   &\multicolumn{1}{c}{$\beta_{E2}$} \\
      $i$  &    $f$     & DHF               & DHF+RPA & +$\Sigma$
                                                                   & DHF+RPA  \\
\hline
$5f_{5/2}$ & $6d_{3/2}$ & 0.015             & 0.0037  & 0.015      & 2 $\times 10^8$  \\
           & $6d_{5/2}$ & 0.0015            & 0.051   & 0.060      & 5 $\times 10^7$ \\
           & $7s_{1/2}$ & 2$\times 10^{-9}$ & 0.032   & 0.037      & 5 $\times 10^7$  \\[2mm]
$7s_{1/2}$ & $7p_{1/2}$ & 18                & 19      & 23         & 7 $\times 10^8$   \\
           & $7p_{3/2}$ & 4.1               & 4.4     & 5.3        & 1 $\times 10^8$  \\
\end{tabular}
\end{ruledtabular}
\end{table}
Including the RPA corrections is formally reduced to replacement of the ``bare''
operators by the ``dressed'' operators. In particular, solving the inhomogeneous
equation we have to replace the operators $\mathcal{T}_K$ in \eref{G2new}
and $D$ in \eref{psif} by $(\mathcal{T}_K)_{\rm eff}$ and $D_{\rm eff}$,
correspondingly.

As seen from the table in certain cases the inclusion of the RPA corrections
increases the probability of the EB process by several orders of magnitude.
It happens, for example, for the $5f_{5/2} \rightarrow 7s$ transition. The channel
$5f_{5/2} \stackrel{\mathcal{T}_1}{\longrightarrow} n \stackrel{E1}{\longrightarrow} 7s$
turns out strongly enhanced because the ``dressed'' MEs
$\langle 5f_{5/2} ||(\mathcal{T}_1)_{\rm eff}|| n \rangle$
are much larger in absolute value than the ``bare'' MEs $\langle 5f_{5/2} ||\mathcal{T}_1|| n \rangle$.
Indeed, we have to consider the intermediate states $n$ that admit
the $E1$ transitions $n \rightarrow 7s$. But for such $n$ the ``bare'' MEs
$|\langle 5f_{5/2} ||\mathcal{T}_1|| n \rangle|$ are very small.

The coefficient $\beta_{M1}$ is rather small for the
$5f_{5/2} \rightarrow 6d_{3/2}$ transition and the RPA and the $\Sigma$ corrections
change its value significantly. The reason is that $G_1^{(1)}$,  $G_{12}^{(1)}$,
and  $G_2^{(1)}$ are comparable in their magnitudes but $G_{12}^{(1)}$
is negative. In the DHF+RPA approximation
it leads to a large cancellation between these terms.

When we consider the $7s$ state as the initial state, the main channel of the process is
$7s \stackrel{\mathcal{T}_1}{\longrightarrow} 8s \stackrel{E1}{\longrightarrow} 7p_{1/2,3/2}$.
Respectively, the first diagram in Fig.~\ref{Fig:EB} (the term $G_1^{(1)}$) gives
the main contribution to $\Gamma_{\rm EB}$ while $G_{12}^{(1)}$ and $G_2^{(1)}$ only slightly
correct this value. As is seen
$\beta_{M1}(7s \rightarrow 7p_j)$ are 2-3 orders of magnitude larger than
$\beta_{M1}(5f_{5/2} \rightarrow 6d_j;7s)$. This is due to the large value of
the ME $\langle 7s ||\mathcal{T}_1|| 8s \rangle$.

As is seen from \tref{betaEB}, the inclusion of the core-valence
correlations changes the values of $\beta_{M1}$ at the level of 20\%
for all considered transitions except the $5f_{5/2} \rightarrow 6d_{3/2}$ transition.
These corrections are not too large because the core orbitals lay rather deep.
In particular the single-electron energy of the external core $6p_{3/2}$ orbital is $-2.1$ au.
For the same reason the contribution to $\Gamma_{\rm EB}$
from the core electrons excitations is small. It is at the level of few per cent.

We also present in \tref{betaEB} the coefficients $\beta_{E2}$ obtained
for the $5f_{5/2} \rightarrow 6d_j;7s$ and the $7s \rightarrow 7p_j$ transitions
in the DHF+RPA approximation. We restricted ourselves by this simple approximation
because these values are given mostly for reference and an order of magnitude
estimate of these quantities is sufficient.

As we have already mentioned above the coefficients $\beta_{E2}$ are many orders
of magnitude larger than the coefficients
$\beta_{M1}$ found for the same transitions. In particular, for the $5f_{5/2} \rightarrow 6d_{3/2}$
transition $\beta_{E2}/\beta_{M1} \sim 10^{10}$.
It is not surprising if we take into account the small value of $k_N^2$ in the
denominator of \eref{betaE2} and the resonant character of the
$5f_{5/2} \stackrel{\mathcal{T}_2}{\longrightarrow} 7p_{1/2}
\stackrel{E1}{\longrightarrow} 6d_{3/2}$ transition because the frequency of
the $5f_{5/2}$ -- $7p_{1/2}$ transition
$\omega_{7p_{1/2},5f_{5/2}} \approx 7.5 \,\, {\rm eV}$ is very close
to $\omega_N = 7.6 \,\, {\rm eV}$.

In spite of that the main contribution to $\Gamma_{\rm EB}$ comes from the
$i \stackrel{\mathcal{T}_1}{\longrightarrow} n \stackrel{E1}{\longrightarrow} f$
channel. As it follows from the results listed in \tref{betaEB} and \eref{M1E2},
we can neglect the contribution to $\Gamma_{\rm EB}$ coming from the
$i \stackrel{\mathcal{T}_2}{\longrightarrow} n \stackrel{E1}{\longrightarrow} f$
channel and put $\Gamma_{\rm EB} \approx \Gamma^{(1)}_{\rm EB}$.

Using Eqs.~(\ref{Gam_N}) and (\ref{B_M1})
we find $\Gamma_N(M1) \approx 6.6 \times 10^{-4} \,\, {\rm sec}^{-1}$ at
$\omega_N = 7.6$ eV and, correspondingly,
\begin{equation}
\Gamma_{\rm EB} \approx
\Gamma^{(1)}_{\rm EB} \approx 6.6 \times 10^{-4} \, \beta_{M1} \,\, {\rm sec}^{-1} .
\end{equation}
The numerical results obtained for $\Gamma^{(1)}_{\rm EB}$
with use of the equation written above are listed in \tref{Gam_EB}.
\begin{table}
\caption{The probabilities $\Gamma^{(1)}_{\rm EB}$
(in sec$^{-1}$) obtained for certain $i \rightarrow f$ transitions for $\omega_N = 7.6$ eV
in the DHF+RPA+$\Sigma$ approximation are presented.}
\label{Gam_EB}
\begin{ruledtabular}
\begin{tabular}{ccc}
      $i$  &    $f$     & $\Gamma^{(1)}_{\rm EB}$ \\
\hline
$5f_{5/2}$ & $6d_{3/2}$ &   9.9 $\times 10^{-6}$    \\
           & $6d_{5/2}$ &   4.0 $\times 10^{-5}$    \\
           & $7s_{1/2}$ &   2.4 $\times 10^{-5}$   \\[2mm]
$7s_{1/2}$ & $7p_{1/2}$ &   1.5 $\times 10^{-2}$      \\
           & $7p_{3/2}$ &   3.5 $\times 10^{-3}$    \\
\end{tabular}
\end{ruledtabular}
\end{table}

\section{Conclusion}
\label{sec_C}
In conclusion, we have calculated the ratios of the probabilities $\Gamma^{(1)}_{\rm EB}$
and $\Gamma^{(2)}_{\rm EB}$ to the probabilities of the nuclear radiative $M1$ and $E2$ transitions,
$\beta_{M1}$ and $\beta_{E2}$. We found that if the valence electron is in
the ground state the coefficients $\beta_{M1}$ are rather small for
all considered transitions.
If the valence electron is in the metastable $7s$ state
the coefficients $\beta_{M1}$ are 2-3 orders of magnitude larger and
$\Gamma_{\rm EB}/\Gamma_N(M1) \sim 20$.

The spectrum of Th$^{3+}$ is not too dense. As a result for the
$i \stackrel{\mathcal{T}_1}{\longrightarrow} n \stackrel{E1}{\longrightarrow} f$
transitions considered in this work there are no electronic transitions which would be at
resonance with the nuclear transition at $\omega_N = 7.6 \,\, {\rm eV}$.

We have found the coefficients $\beta_{E2}$ to be many orders of
magnitude larger than $\beta_{M1}$, but based upon the estimate
$\Gamma_N(M1)/\Gamma_N(E2) \sim 10^{11}$ one can state that the contribution
of the $i \stackrel{\mathcal{T}_2}{\longrightarrow} n \stackrel{E1}{\longrightarrow} f$
channel to $\Gamma_{\rm EB}$ is negligible. It is worth noting that this statement
is correct for all considered transitions in spite of the resonant character of the
$5f_{5/2} \stackrel{\mathcal{T}_2}{\longrightarrow} 7p_{1/2}
\stackrel{E1}{\longrightarrow} 6d_{3/2}$ transition.

We would like to thank J.~Berengut and A.~Kuzmich for stimulating
discussion. This work was supported by Australian Research Council.
The work of S.G.P. was supported in part by the Russian Foundation for Basic
Research under Grants No. 07-02-00210-a and No. 08-02-00460-a.

\appendix
\section{}
\label{Ap}
The expressions for the single-electron operators $T_1$ and $T_2$ can
be written as
\begin{equation}
T_{1 \lambda}({\bf r}) =
\frac{-i \sqrt{2}\, {\bm \alpha} \cdot {\bf C}^{(0)}_{1 \lambda}({\bf n})}{c\,r^2}
\end{equation}
and
\begin{equation}
T_{2 \lambda}({\bf r}) =
\frac{-C_{2 \lambda}({\bf n})}{r^3} ,
\end{equation}
where ${\bf n} \equiv {\bf r}/r$ and ${\bf C}^{(0)}_{K \lambda}$ is a normalized vector spherical harmonic
defined by (see, e.g.,~\cite{VarMosKhe88})
\begin{equation}
{\bf C}^{(0)}_{K \lambda}({\bf n}) = \frac{{\bf L}} {\sqrt{K(K+1)}} C_{K \lambda}({\bf n}) .
\end{equation}
Here ${\bf L}$ is the orbital angular momentum operator and
$C_{K \lambda}$ is a spherical harmonic given by
\begin{equation}
 C_{K \lambda}({\bf n}) = \sqrt{\frac{4\pi}{2K+1}}\, Y_{K \lambda}({\bf n}) .
\end{equation}
To calculate the MEs of the operators $D$, $T_1$, and $T_2$ we define the one-electron
wave function $|a\rangle \equiv \psi_a({\bf r})$ as follows
\begin{equation}
\psi_a ({\bf r}) = \frac{1}{r}
\left(
\begin{array}{l}
P_a(r)\,\, \Omega_{ \kappa_a m_a} ({\bf n}) \\
i Q_a(r)\, \Omega_{-\kappa_a m_a}({\bf n})
\end{array}
\right) ,
\end{equation}
where $\kappa_a = (l_a-j_a)(2j_a+1)$.

Using the ME $\langle \kappa_b ||C_K|| \kappa_a \rangle$ :
\begin{eqnarray*}
\langle \kappa_b ||C_K|| \kappa_a \rangle &=&
(-1)^{j_b+1/2} \sqrt{(2j_a+1)(2j_b+1)} \nonumber  \\
&\times&\!  \left( \!
\begin{array}{ccc}
 j_b & j_a & K \\
-1/2 & 1/2 & 0
\end{array}
\! \right) \xi(l_b+l_a+K) ,
\end{eqnarray*}
where
$$ \xi(x) =
\left\lbrace
\begin{array}{l}
1,\, {\rm if}\,\, x\,\, {\rm is\,\, even}  \\
0,\, {\rm if}\,\, x\,\, {\rm is\,\, odd} ,
\end{array}
\right. $$
\noindent
we can write the reduced ME for the electric dipole operator $D$
in the following form
\begin{eqnarray}
&&\langle n_b \kappa_b ||D|| n_a \kappa_a \rangle =
  \langle n_b \kappa_b ||-r|| n_a \kappa_a \rangle = \nonumber \\
&& -\langle \kappa_b ||C_1|| \kappa_a \rangle
 \int_0^\infty \! \left\{P_b Q_a  + \! P_a Q_b \! \right\} r \, dr ,
\end{eqnarray}
where $n_i$ is the principal quantum number.

The reduced ME for the magnetic dipole operator $T_1$ is represented by
\begin{eqnarray}
&&\langle n_b \kappa_b ||T_1|| n_a \kappa_a \rangle =
 \langle -\kappa_b ||C_1|| \kappa_a \rangle  \nonumber  \\
&& \times (\kappa_b + \kappa_a) \int_0^\infty \!
\left\{P_b Q_a  + \! P_a Q_b \! \right\} \frac{1}{r^2}\, dr .
\label{Ap:T1}
\end{eqnarray}
Rewriting the angular part of \eref{Ap:T1} in a more simple form we arrive at
\begin{eqnarray}
&&\langle n_b \kappa_b ||T_1|| n_a \kappa_a \rangle =
\xi(l_b+l_a) (-1)^{j_a+l_a-1/2} \nonumber  \\
&&\times \frac{c_{j_a j_b}}{2}
\int_0^\infty \! \left\{P_b Q_a  + \! P_a Q_b \! \right\} \frac{1}{r^2}\, dr ,
\label{T1}
\end{eqnarray}
where
\noindent
$ c_{j_a j_b}\! \equiv \!
\left\lbrace
\begin{array}{l}
\sqrt{ (2j_a+1)(2j_b+1)/(j_{\rm min}+1) },\, j_a \neq j_b   \\
\sqrt{ (2j_a+1)^3/(j_a(j_a+1)) }, \, j_a = j_b
\end{array}
\right. $
and $j_{\rm min} = {\rm min} \,(j_a, j_b)$.

The reduced ME for the electric quadrupole operator $T_2$ is given by
\begin{eqnarray}
\langle n_b \kappa_b ||T_2|| n_a \kappa_a \rangle &=&
-\langle \kappa_b ||C_2|| \kappa_a \rangle  \nonumber  \\
&\times& \int_0^\infty \!
\left\{P_a P_b  + \! Q_a Q_b \! \right\} \frac{1}{r^3}\, dr .
\label{Ap:T2}
\end{eqnarray}


\end{document}